\documentclass{article}
\usepackage{url,epsfig}
\usepackage{amsmath}
\usepackage{latexsym}
\usepackage{amsfonts}
\usepackage{amsmath}
\usepackage{epsf,amsfonts,amssymb}
\usepackage{amssymb}
\usepackage{amsmath}

\begin{document}

\title{A new model for the immune clonal networks }
\author{V. V. Gafiychuk\thanks{Institute of computer modeling,
Krakow University of Technology, 24 Warszawska Street, 31155,
Krakow, Poland; Institute of Applied Problems of
Mechanics and Mathematics NASU, Lviv, Ukraine, 79053.} \ \ and A.K. Prykarpatsky\thanks{%
Dept. of Applied Mathematics at the AGH University of Science and
Technology, 30 Mickiewicz Al. Bl. A4, 30059 Krakow, Poland;
Institute of Applied Problems of Mechanics and Mathematics NASU,
Lviv, Ukraine, 79053.}} \maketitle

\begin{abstract}
This paper deals with a new model for clonal network dynamics. We describe
in detail this model and derive special equations governing immune system
dynamics based on the general gradient type principles that can be inherent
to a wide class of real living objects. A special clonal network is modeled
by two symmetric projector matrix variables simultaneously taking into
account both asymmetry of the interaction to each other and adaptation
states that can be realized owing to possible idiotypic clonal suppresions.
We perform computer simulations of the model dynamics for some simple cases
of relatively low dimension, paying special attention to the dynamics of \
amounts of activated receptor strings within clonal network.

\noindent \textbf{keywords}: clonal dynamics, immune system, mathematical
model, gradient dynamical system, Lyapunov function, computer simulation,
complex system.
\end{abstract}

\section{Introduction}

The dynamics of clonal network is a very important problem for understanding
immune systems dynamics and starting from the first mathematical model of
Jerne \cite{jer} a lot of mathematical models are used for modeling it (See,
for example \cite{4a,ct02,ks00,seg,db,mar}). It is very well known that
immune system contains many types of B-cells which can activate each other
at certain conditions. If the receptors of such B-cells match each other the
cells become antibody cells and they effectively recognize antigens \cite%
{4a,seg,db,mar}. The construction of the cell gene model allows one to
understand how the same network can support different independent state of
the immune system. Among these states we can distinguish states which are
quite different in their functional activity, for example a virgin state
when the clone state is not activated, the immune state when a clone
proliferates antibody cells and finally - the inhibited state \ when
proliferation terminates. Which state is reached depends on the local
topology of the network and conditions of gene presentation With the
additional assumption that the parameters of the model change between early
and late states one can understand how self-nonself interaction is
accomplished by the clonal network. In this case several original approaches
has been proposed in order to explain characteristic features of evolution
of B-cell repertoire \cite{noe,leg} and evolution of specificity in humoral
immunity based on idiotypic network \cite{lui}.

Dynamics of this clonal type network with symmetric interactions probably
can converge \cite{2} to steady or oscillatory state under very general
assumptions \cite{and}. But in reality interactions between clones are not
confined to steady states (attractors) but also include other almost
periodic and aperiodic behaviors. The interaction between an excitatory and
inhibitory clone is clearly asymmetric. Assume that a self basic gene is
seen by its own clone. Recall \cite{3} that adaptation can be achieved by
the suppression of this idiotypic clone by the corresponding anti-idiotypic
clone. Once the adaptation state is achieved, secondary presentation of the
leading gene does not give rise to any network response, because the gene
only increases the suppression on its clone. It would be interesting to
observe if some basic modifications to the governing network model equations
could be made that allow really memory retention \cite{1,par}. If it is not
the case, then not being able to retain memory of former inhibitory gene
encounters may not be failing of the model under regard but rather a
reflection of the intrinsic properties of clonal network.

The model is composed \cite{4a,4,5,7,8,9,10} of a varying number of cell
clones of different specificities that form a clonal network. Each clone is
characterized by its specific inhibitory and excitatory receptors, which are
specified in the model by bit strings and clonal receptor string $x_{\alpha
}\in \mathbb{R}^{n},\alpha \in \mathcal{N}_{x}$ and $y_{\beta }\in \mathbb{R}%
^{n},$ $\beta \in \mathcal{N}_{y},$ normalized by the conditions $\left\Vert
x_{\alpha }\right\Vert =1$ and $\left\Vert y_{\beta }\right\Vert =1$
allowing for some components to bring either positive or negative signs.
Thereby this makes it possible to achieve suitable adaptation state through
the related suppression of idiotypic clone by the corresponding
anti-idiotypic clone, whose modeling within our clonal network \ is realized
in the framework of the postulated self-similarity of gene action and
asymmetry of clone interaction. Two clones can interact via soluble
inhibitory genes whenever their receptor shapes, i.e. bit strings are
complementary. Cells that become activated proliferate and differentiate
into inhibitory secreting cells. This process takes some time during which
another free inhibitory cells form dynamic complexes not taking often real
role within excitatory-inhibition network dynamics.

The existence of invariant localized dynamic patterns says that our
idiotypic clonal network possesses certain self-structured properties.
Asymmetric interactions within the network determinate network's working
size and connectivity and determine the total inhibitory cells level.
Concerning the connectivity within the network one can conclude that the
network in equilibrium selects for growth and retaining within the network
the clones with low connectivity. Within this framework low connectivity is
not an intrinsic property of any particular inhibitory cell but rather is
determined by the random structure of a clone's receptor and the shapes of
the receptors on the other clones presented in the system.

The incorporation of this meta-dynamics in this excitatory-inhibitory model
one can consider as an attempt to account for the rapid enough turn over of
clones in the clonal network. Such processes were also studied in \cite%
{8,9,cel,ban,ch} within a cellular automation approach. There was found that
networks of automata can be considered as dynamical systems being the
discrete equivalent of differential systems. They have been recently widely
used in clonal nets and cellular automata to model complex systems such as
brain. One of the main advantages of these networks for a biologist involved
in modeling is that the construction of a model requires a minimal knowledge
about the numerical values of the parameters defining a system. The
differential equation systems describing require biological data on cell
lifetimes like thresholds for activation, affinity constraints of the
network dynamics. Most automata models do not require these data, since the
basic assumption is that sell populations need only be described by a set of
some discrete values, often 0 and 1, where 0 means that a populations
absent, while 1 means that it is presented at a high enough level. The
corresponding interactions among populations are represented by logical
functions, i.e. Boolean set function, which most often are equivalent to
threshold automata like physical spin systems \cite{8,9,cel,ban}. Such a
modeling can be applied to a wide class of complicated neural networks
taking into account different states of development of the corresponding
cells and also to system evolution differential equations discussed below.
These aspects of studying our excitatory-inhibitory clonal type network are
planned to be discussed in more detail in another place. In this paper we
will use in part an approach devised in for \cite{gaf1,gaf2} describing
evolution of the introduced clonal network.

\section{Clonal network model description}

A network under consideration models a clonal dynamics exhibiting
excitatory-inhibitory properties. It consists of interacting cell clones
generated by inhibitory and excitatory genes within a fixed medium. The
latter will be called a configuration phase space, depending strongly on the
nature of interaction between clones. An inhibitory clone population within
the network can be effectively encoded, in general, by a real $(n\times n)-$%
\ asymmetric projector matrix of the canonical \cite{11} form $X:=\overset{%
n(x)}{\underset{\alpha \in \mathcal{N}_{x}}{\sum }}x_{\alpha }\otimes \hat{x}%
_{\alpha }\in End\mathbb{R}^{n},$\ where biorthogonal to each other vectors $%
x_{\alpha },\hat{x}_{\alpha }\in \mathbb{R}^{n},$ $\ \alpha \in \mathcal{N}%
_{x}\subset \overline{1,n},$\ are the corresponding reciprocity receptor
strings amplitudes responsible for an inhibitory clonal topology within the
network and the number $card\mathcal{N}_{x}=$ $n(x)\in \mathbb{Z}_{+}$\ \
means exactly the amount of activated receptors belonging \ to the
inhibitory clone. The inhibitory clonal self-similarity is realized now by
means of the fundamental projector property $X\cdot X=X$ for all whiles of
time.

Similarly, an excitatory population can be encoded, in general, by real $%
(m\times m)$ -asymmetric projector matrix of the canonical form \ $Y:=%
\overset{n(y)}{\underset{\beta \in \mathcal{N}_{y}}{\sum }}y_{\beta }\otimes
\hat{y}_{\beta }\in End\mathbb{R}^{m},$ where biorthogonal to each other
excitatory vectors $y_{\beta },\hat{y}_{\beta }\in \mathbb{R}^{m},$ $\beta
\in \mathcal{N}_{y}\subset \overline{1,m},$\ \ are the corresponding
reciprocity receptor strings amplitudes responsible for an excitatory clonal
receptor topology and $card\mathcal{N}_{y}=n(y)\in \mathbb{Z}_{+}$\ means
the amount of activated excitatory receptors during the network dynamics.
The excitatory clonal self-similarity is realized here also by means of the
fundamental projector property $Y\cdot Y=Y$ for all whiles of time. In
general, the integer numbers $n(x),n(y)\in \mathbb{Z}_{+}$ can change during
the network dynamics because of \ the possible full suppression of some
activated clone receptor strings. Concerning the interaction between the
clonal populations, it is described by means of a real $(m\times n)$\
-matrix $Z:=\overset{n(x)}{\underset{\alpha \in \mathcal{N}_{x}}{\sum }}%
\overset{n(y)}{\underset{\beta \in \mathcal{N}_{y}}{\sum }}z_{\beta \alpha
}y_{\beta }\otimes \hat{x}_{\alpha }\in Hom(\mathbb{R}^{n};\mathbb{R}^{m})$\
with parameters $z_{\beta \alpha }\in \mathbb{R},$\ $\alpha \in \mathcal{N}%
_{x}\subset \overline{1,n},$ $\beta \in \mathcal{N}_{y}\subset \overline{1,m}%
,$\ responsible for the strengths of interaction between receptors strings
of two clonal populations. The adaptation of some inhibitory or excitatory
during interaction clones is modeled within our clonal network by means of a
possible time dependence of corresponding strength parameters $z_{\beta
\alpha }\in \mathbb{R}$, $\ \alpha \in \mathcal{N}_{x}\subset \overline{1,n}%
, $ $\beta \in \mathcal{N}_{y}\subset \overline{1,m},$ \ that the matrix $Z$%
\ necessarily satisfies the next important self-similar clonal interaction
properties $ZX=Z=YZ$ for all whiles of time.

\section{Clonal network topology and dynamics}

It is natural to endow our configuration phase space $M_{(x,y,z)}\subset (End%
\mathbb{R}^{n}{\footnotesize \times }End\mathbb{R}^{m})$ \ $\times $ $Hom(%
\mathbb{R}^{n};\mathbb{R}^{m})$ with a reasonable Riemannian metrics by
means of the following scalar product on its tangent space $T(M):$
\begin{equation}
\left\langle (X,Y,Z),(\widetilde{X},\widetilde{Y},\widetilde{Z}%
)\right\rangle :=tr(X^{T}\widetilde{X})+tr(Y^{T}\widetilde{Y})+tr(Z^{T}%
\widetilde{Z})  \label{1}
\end{equation}%
where $\ X,\widetilde{X}\in T(End\mathbb{R}^{n}),$ $Y,\widetilde{Y}\in T(End%
\mathbb{R}^{m})$ \ and $Z,\tilde{Z}\in T(Hom(\mathbb{R}^{n};\mathbb{R}^{m}))$
are arbitrary elements of the corresponding tangent spaces. Concerning \ the
metrics (\ref{1}) one can construct a gradient vector field on the projector
field manifold $M_{(x,y,z)}$\ generated by the Lyapunov interaction function
$\Phi :M\rightarrow \mathbb{R}$\ whose variation is%
\begin{equation}
\delta \Phi (X,Y,Z):=tr(D_{h}^{T}\delta Z)+tr(D_{f}^{T}\delta
X)+tr(D_{g}^{T}\delta Y)  \label{2}
\end{equation}%
for some specified matrices $D_{f}\in T(End\mathbb{R}^{n}),$ $D_{g}\in T(End%
\mathbb{R}^{m})$ \ and $D_{h}\in T(Hom(\mathbb{R}^{n};\mathbb{R}^{m})),$
being responsible for the asymmetry of the interaction between clonal
populations. Except the Lyapunov function variation (\ref{2}) it is
necessary to involve into the picture the following natural clonal phase
constraints:

\begin{eqnarray}
tr(A^{T}(X^{2}-X)) &=&0,\ tr(B^{\intercal }(Y^{2}-Y))=0,  \notag \\
tr((ZX-Z)Q^{\intercal }) &=&0,\ \ tr((YZ-Z),R^{T})=0,  \label{3}
\end{eqnarray}%
holding for any $A\in End\mathbb{R}^{n},B\in End\mathbb{R}^{m}$\ and $Q,R\in
Hom(\mathbb{R}^{n};\mathbb{R}^{m}).$\ Constraints (\ref{3}) can be still
augmented in many special cases by the symmetry conditions%
\begin{equation}
tr((X^{\intercal }-X)P^{\intercal })=0,\ \ tr((Y^{\intercal }-Y)S^{\intercal
})=0,  \label{4}
\end{equation}%
holding also for arbitrary matrices $P\in End\mathbb{R}^{n}$ and $S\in End%
\mathbb{R}^{m}.$

Below we will consider only this strongly symmetric case of our clonal
network. Concerning the constraint conditions involved above the
corresponding gradient vector field generated by the Lyapunov function
variation (\ref{2}) is given as%
\begin{eqnarray}
dX/dt &=&[[D_{f},X],X]+  \notag \\
&&+(Z^{\intercal }Z+2I)^{-1}(Z^{\intercal }D_{h}-Z^{\intercal }ZD_{f})(1-X)+
\notag \\
&&+(1-X)(D_{h}^{\intercal }Z-D_{f}Z^{\intercal }Z)(Z^{\intercal }Z+2I)^{-1},
\notag \\
dY/dt &=&[[D_{g},Y],Y]+  \notag \\
&&+(1-Y)(D_{h}Z^{\intercal }-D_{g}ZZ^{\intercal })(ZZ^{\intercal }+2I)^{-1}+
\notag \\
&&+(2I+ZZ^{\intercal })^{-1}(ZD_{h}^{\intercal }-ZZ^{\intercal }D_{g})(1-Y),
\notag \\
dZ/dt &=&-D_{h}X-ZD_{f}X+  \notag \\
&&+2(Y-1)(D_{h}X-D_{g}Z)(Z^{T}Z+2I)^{-1}+  \notag \\
&&+Z(Z^{T}Z+2I)^{-1}(Z^{\intercal }D_{h}-Z^{\intercal }ZD_{f})(1-X)
\label{5}
\end{eqnarray}%
with the canonical \cite{11} representations $\ $ $X:=\overset{n(x)}{%
\underset{\alpha \in \mathcal{N}_{x}}{\sum }}x_{\alpha }\otimes x_{\alpha
}\in End\mathbb{R}^{n},$ $Y:=\overset{n(y)}{\underset{\beta \in \mathcal{N}%
_{y}}{\sum }}y_{\beta }\otimes y_{\beta }\in End\mathbb{R}^{n}$, $%
\left\langle x_{\alpha },x_{\alpha ^{\prime }}\right\rangle =\delta _{\alpha
\alpha ^{\prime }},$ $\alpha ,\alpha ^{\prime }\in \mathcal{N}_{x}\subset
\overline{1,n},$ $\left\langle y_{\beta },y_{\beta ^{\prime }}\right\rangle
=\delta _{\beta \beta ^{\prime }},$ $\beta ,\beta ^{\prime }\in \mathcal{N}%
_{y}\subset \overline{1,m}.$ Here we put also $D_{f}=D_{f}^{\intercal },$ $%
D_{g}=D_{g}^{\intercal }$ and took the matrix $D_{h}\in T(End(\mathbb{R}^{n};%
\mathbb{R}^{m}))$ arbitrary.

The gradient vector field (\ref{5}) \ can be more specified if to take into
account the following Lyapunov function special case taking into account the
corresponding self-symmetry clonal declinations:
\begin{equation}
\Phi ^{(1)}=tr(\alpha _{h}(Z-X\beta _{f}Z^{\intercal }\beta _{g}^{\intercal
}Y)+tr(Z^{\intercal }-Y\beta _{g}Z\beta _{f}^{\intercal }X)\alpha
_{h}^{\intercal }),  \label{6}
\end{equation}%
where $(X,Y,Z)\in M_{(x,y,z)}$\ and $\beta _{f}\in End\mathbb{R}^{n}$\ and $%
\beta _{g}\in End\mathbb{R}^{m}$\ are some constant matrices which are close
to unity matrices with respect to the corresponding norms in $End\mathbb{R}%
^{n}$\ and $End\mathbb{R}^{m}$. Then owing to the definition (\ref{2}), one
finds that%
\begin{eqnarray}
D_{h}^{(1)} &=&2(\alpha _{h}-\beta _{g}^{\intercal }Y\alpha _{h}X),\   \notag
\\
D_{f}^{(1)} &=&-(\beta _{f}Z^{\intercal }\beta _{g}^{\intercal }Y\alpha
_{h}+\alpha _{h}^{\intercal }Y\beta Z\beta _{f}^{\intercal }),\   \notag \\
D_{g}^{(1)} &=&-(\alpha _{h}X\beta _{f}Z^{\intercal }\beta _{g}^{\intercal
}+\beta _{g}Z\beta _{f}^{\intercal }Z\alpha _{h}^{\intercal }),  \label{7}
\end{eqnarray}%
Similarly one can take the following Lyapunov type function

\begin{eqnarray}
\Phi ^{(2)} &=&tr(\alpha _{h,f}(Z^{\intercal }-X\beta _{f}Z^{\intercal
}))+tr(\alpha _{h,g}(Z^{\intercal }-Z^{\intercal }\beta _{g}Y))+  \notag \\
&&tr((Z-Z\beta _{f}^{\intercal }X)\alpha _{h,f}^{\intercal })+tr((Z-Y\beta
_{g}^{\intercal }Z)\alpha _{h,g}^{\intercal }),  \label{8}
\end{eqnarray}%
where $(X,Y,Z)\in M_{(x,y,z)},$\ $\beta _{f}\in End\mathbb{R}^{n}$\ and $%
\beta _{g}\in End\mathbb{R}^{m}$ are some constant matrices which are close
to unity matrices with respect to the corresponding norms in $End\mathbb{R}%
^{n}$\ and $End\mathbb{R}^{m}$. Then owing to the definition (\ref{2}), one
finds easily that%
\begin{eqnarray}
D_{h}^{(2)} &=&2\alpha _{h,f}(I-X\beta _{f})+2(I-\beta _{g}Y)\alpha _{h,g},
\notag \\
D_{f}^{(2)} &=&-(\beta _{f}Z^{\intercal }\alpha _{h,f}+\alpha
_{h,f}^{\intercal }Z\beta _{f}^{\intercal }),  \notag \\
D_{g}^{(2)} &=&-(\beta _{g}^{\intercal }Z\alpha _{h,g}^{\intercal }+\alpha
_{h,g}Z^{\intercal }\beta _{g}).  \label{8a}
\end{eqnarray}%
It has to be mentioned here that the Lyapunov function (\ref{6}) models in
general an asymmetric case of the mutual interaction between inhibitory and
excitatory clone populations, realizing a really observed pattern formation
structure, when the whole system is both under an external medication and
intrinsically activated immune state \cite{cal}.

The special analysis still must be done concerning the possible similarities
between receptor sets of inhibitory and excitatory genes. This means that
some mutual relationship between projector clone operators can be realized.
For instance, their complete weak orthogonality can be realized and the
following additional constraint
\begin{equation}
tr(Y\xi X\eta ^{T})=0  \label{9}
\end{equation}%
for all \ $(X,Y,Z)\in M_{(x,y,z)}$ and some constant matrices $\xi ,\eta \in
Hom(\mathbb{R}^{n};\mathbb{R}^{m})$ can hold$.$ The condition (\ref{9})
gives rise to a little complicated gradient field like (\ref{5}) in part
reflecting the mentioned above pattern isolating property of our clonal
populations.

It can be also interesting to analyze the dynamical system (\ref{8}) in the
case when either projector matrices $X\in End\mathbb{R}^{n}$ or $Y\in End%
\mathbb{R}^{m}$ or both ones model clonal populations with the fixed number
of the corresponding receptor strings activated during the interaction
between them be realized that can happen when the clonal network is
activated artificially by means of some external medication. This means that
integers $trX=n(x)\in \mathbb{Z}_{+}$ or $trY=n(y)\in \mathbb{Z}_{+}$\
persist to be fixed during the system evolution. Then these integers must be
conserved quantities\ \ for all $t\in \mathbb{R}$\ involving the additional
scalar constraints. The latter gives rise to a little modified gradient
dynamical system like (\ref{5})\ which we do not write down here, being
obtained the same way as before.

\section{Spectral analysis}

Consider now two respectively biorthogonal systems of vectors $x_{\alpha
}\in \mathbb{R}^{n},$\ $\alpha \in \mathcal{N}_{x}\subset \overline{1,n},$\
and $y_{\beta }\in \mathbb{R}^{m},\beta \in \mathcal{N}_{y}\subset \overline{%
1,m},$\ being eigenvectors of the symmetric projector \ matrices $%
X=X^{\intercal }\in End\mathbb{R}^{n}$ and $Y=Y^{\intercal }\in End\mathbb{R}%
^{m},$ respectively, satisfying \cite{11} the following conditions:

\begin{equation}
Xx_{\alpha }=x_{\alpha }\,,\ \text{ }\ Yy_{\beta }=y_{\beta }\text{ }
\label{10}
\end{equation}%
for all \ $\alpha \in \mathcal{N}_{x}\subset \overline{1,n}$\ and \ $\beta
\in \mathcal{N}_{y}\subset \overline{1,m}.$ \ By differentiating the
equalities (\ref{10}) with respect to the time, one gets that%
\begin{equation}
dX/dt\text{ }x_{\alpha }+Xdx_{\alpha }/dt=dx_{\alpha }/dt,\,\ \ dY/dt\text{ }%
y_{\beta }+Ydy_{\beta }/dt=dy_{\beta }/dt  \label{11}
\end{equation}%
for all $\alpha \in \mathcal{N}_{x}\subset \overline{1,n}$\ and \ $\beta \in
\mathcal{N}_{y}\subset \overline{1,m}.$ Making use of the equations (\ref{5}%
) and (\ref{10}), one has that%
\begin{eqnarray*}
\left[ (I-X)(D_{f}X-XD_{f})+K_{f}(I-X)+(I-X)K_{f}\right] x_{\alpha }
&=&(I-X)dx_{\alpha }/dt, \\
\left[ (I-Y)(D_{g}Y-YD_{g})+K_{g}(I-Y)+(I-Y)K_{g}\right] &=&(I-Y)dy_{\beta
}/dt,
\end{eqnarray*}%
where, by definition, matrices $K_{f}\in End\mathbb{R}^{n},$ $K_{g}$\ $\in
End\mathbb{R}^{m}$ and denote the corresponding out commutative parts of
first two equations in (\ref{5}). As a result of simple computations one
gets that
\begin{eqnarray}
(I-X)(D_{f}X-XD_{f})x_{\alpha }+(I-X)K_{f}x_{\alpha } &=&(I-X)dx_{\alpha
}/dt,  \notag \\
(I-Y)(D_{g}Y-YD_{g})y_{\beta }+(I-Y)K_{g}y_{\beta } &=&(I-Y)dy_{\beta }/dt.
\label{12}
\end{eqnarray}%
From (\ref{12})\ one finds easily that

\begin{eqnarray}
dx_{\alpha }/dt &=&(D_{f}X-XD_{f})x_{\alpha }+(I-X)K_{f}x_{\alpha
}+Xz_{\alpha }^{(f)},  \label{13} \\
dy_{\beta }/dt &=&(D_{g}Y-YD_{g})y_{\beta }+(I-Y)K_{g}y_{\beta }+Yz_{\beta
}^{(g)}  \notag
\end{eqnarray}%
for some vectors $z_{\alpha }^{(f)}\in \mathbb{R}^{n}$\ and $z_{\beta
}^{(g)}\in \mathbb{R}^{m}$ for all $\alpha \in \mathcal{N}_{x}\subset
\overline{1,n}$\ and $\beta \in \mathcal{N}_{y}\subset \overline{1,m}.$
Since, \ in general, $tr(dX/dt)\neq 0$ and $tr(dY/dt)\neq 0,$ we deduce that
the integers $rankX$\ and $rankY$\ are changing in time. On the other hand,
since $trX=n(x)\in \mathbb{Z}_{+}$ and $trY=n(y)\in \mathbb{Z}_{+}$\ are
integers, we see that our the dynamical system possesses very interesting
properties related with jumping of the integers $n(x)$ and $n(y)\in \mathbb{Z%
}_{+}$ at some fixed whiles of time. This phenomenon can be interpreted\ \
naturally as a result of activation (dis-activation) of available receptor
strings characterizing our interacting clonal populations during the
interaction process. Returning \ back to equations (\ref{13}) one can
observe that vectors $z_{\alpha }^{(f)}\in \mathbb{R}^{n}$\ and $z_{\beta
}^{(g)}\in \mathbb{R}^{m}$\ must satisfy the conditions\
\begin{eqnarray}
&<&z_{\alpha }^{(f)},x_{\alpha ^{\prime }}>=0,\text{ \ \ }<z_{\beta
}^{(g)},y_{\beta ^{\prime }}>=0,  \label{14} \\
&<&D_{f}x_{\alpha },x_{\alpha ^{\prime \prime }}>+<K_{f}x_{\alpha
},x_{\alpha ^{\prime \prime }}>=0, \\
\text{\ \ \ \ \ \ \ \ \ \ \ \ \ \ \ \ } &<&D_{g}y_{\beta },y_{\beta ^{\prime
\prime }}>+<K_{g}y_{\beta },y_{\beta ^{\prime \prime }}>=0  \notag
\end{eqnarray}%
for $\alpha ,\alpha ^{\prime }\in \mathcal{N}_{x}\subset \overline{1,n},$ $%
\alpha ^{\prime \prime }\notin $\ $\mathcal{N}_{x}\subset \overline{1,n},$
and for $\beta ,\beta ^{\prime }\in \mathcal{N}_{y}\subset \overline{1,m},$ $%
\beta ^{\prime \prime }\notin \mathcal{N}_{y}\subset \overline{1,m}.$
Relationships (\ref{14}) \ can be used as some criterion for the
corresponding ranks of the projector matrices $\ X\in End\mathbb{R}^{m}$ and
$Y\in End\mathbb{R}^{m},$ evolving with respect to the dynamical system (\ref%
{5}), to become jumped.

\section{Clonal network simulation}

So far introduced mathematical clonal network model has been written as a
formal system of matrix nonlinear differential equations. In order to check
the characteristic features of this system we have performed computer
simulations of the model for several values of $n,m\in \mathbb{Z}_{+}$ and
as well as for different initial conditions and forms of the corresponding
constant matrices entering the system. The model is composed in general of $%
n(n+1)/2+m(m+1)/2+nm\in \mathbb{Z}_{+}$ scalar differential equations, i.e.
of the inhibitor share equations, excitor share equations and equations
describing interaction between these shares. The integration of this system
requires simultaneous integrations of these scalar differential equations,
each of which is a first order ODE. Let us now illustrate the dynamic
behavior of our system by computer simulation at some simple starting data.
We rewrite down governing equations (\ref{5}) in a new equivalent form which
was used for computer simulation:

\begin{eqnarray}
dX/dt &=&[[D_{f},X],X]+  \notag \\
&&+X(XZ^{\intercal }YZ+2I)^{-1}X(Z^{\intercal }YD_{h}-Z^{\intercal
}YZXD_{f})(I-X)+  \notag \\
&&+(I-X)(D_{h}^{\intercal }YZ-D_{f}XZ^{\intercal }YZ)X(XZ^{\intercal
}YZ+2I)^{-1}X,  \notag \\
dY/dt &=&[[D_{g},Y],Y]+  \notag \\
&&+(I-Y)(D_{h}XZ^{\intercal }-D_{g}YZXZ^{\intercal })Y(YZXZ^{\intercal
}+2I)^{-1}Y+  \notag \\
&&Y(2I+YZXZ^{\intercal })^{-1}Y(ZD_{h}^{\intercal }-ZXZ^{\intercal
}YD_{g})(I-Y),  \notag \\
dZ/dt &=&-D_{h}X-YZXD_{f}X+  \notag \\
&&2(Y-I)(D_{h}X-D_{g}YZ)X(XZ^{T}YZ+2I)^{-1}X+  \notag \\
&&+YZX(XZ^{T}YZ+2)^{-1}X(Z^{\intercal }YD_{h}-Z^{\intercal }YZXD_{f})(I-X).
\label{ad1}
\end{eqnarray}

Here when writing down the system (\ref{ad1}) we took into account
constraints (\ref{3}) in order to get matrix paths always lying on the
corresponding projector parts of the manifold $M.$ We used Matlab software
to perform our simulation of equations (\ref{ad1}). Despite the complex
structure of equations describing our immune clonal network, its dynamical
behavior is fairly interesting and good modeling some real ones.

In Fig. 1 we plotted the outcome of numerical simulations for the case $\dim
X=4$, $\dim Y=2$. The plots in Fig. 1 show that the simulations recover
perfectly the clonal network behavior predicted by both qualitative and some
analytic considerations. For simplicity, we first consider the matrix $D_{h}$
being zero. Fig. 1 displays a dynamics of inhibitory network $X$ and
excitatory $Y$ at the conditions and parameters indicated in the capture of
the figure. The behavior of the diagonal elements (receptors mapping antigen
concentration) of the of the excitor are displayed by solid line and
inhibitory receptors are displayed by dashed line. The displayed dynamics
has a simple physical meaning. The receptor of activator network responsible
for antigen state is getting depressed (equal to zero) due to activation of
inhibitory one (Fig. 1 b). At certain value of time the state of the
inhibitory network is tending to zero also.
\begin{figure}[tbph]
\centerline{ \hfill
\psfig{figure=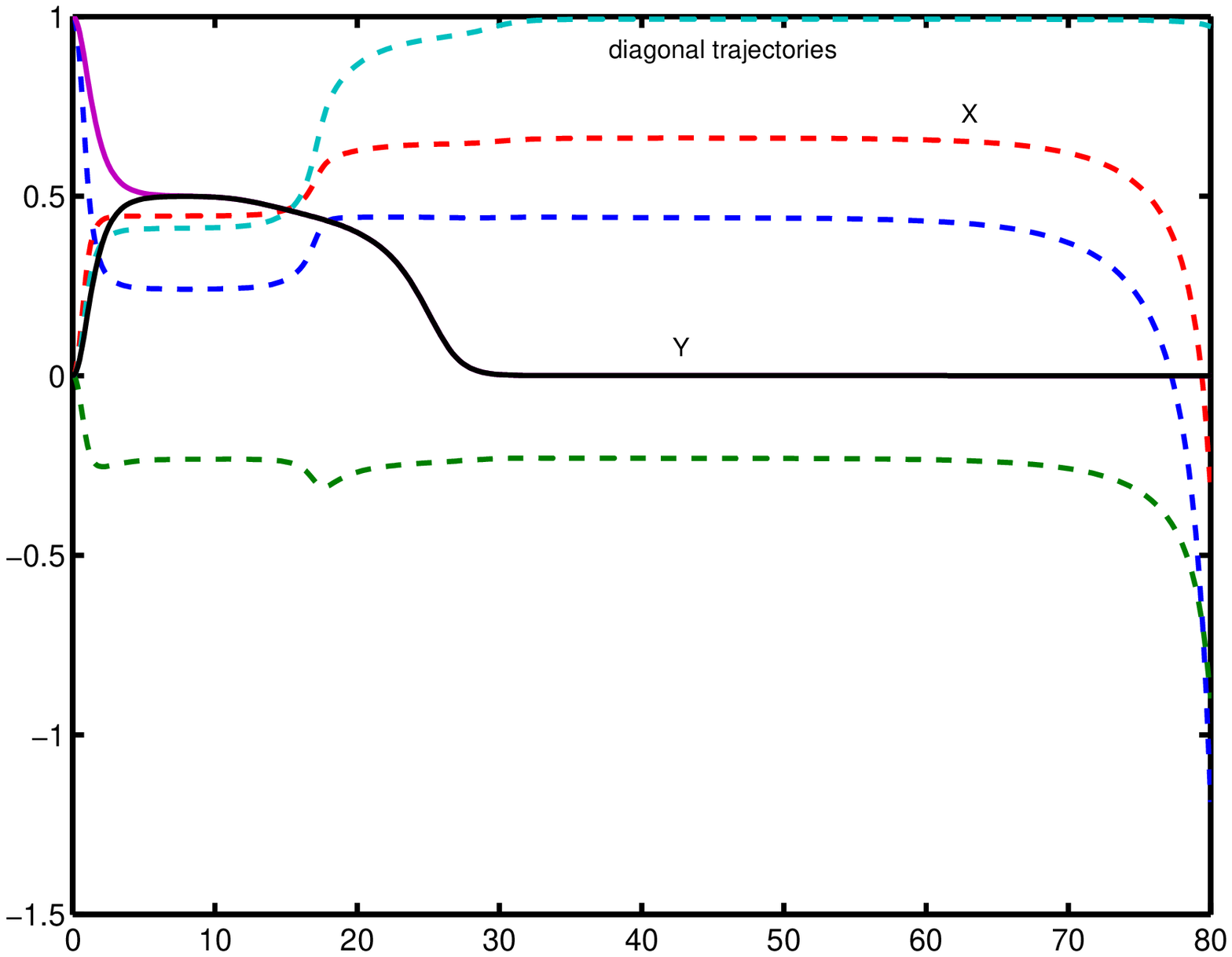,width=6.0cm} \hfill
\psfig{figure=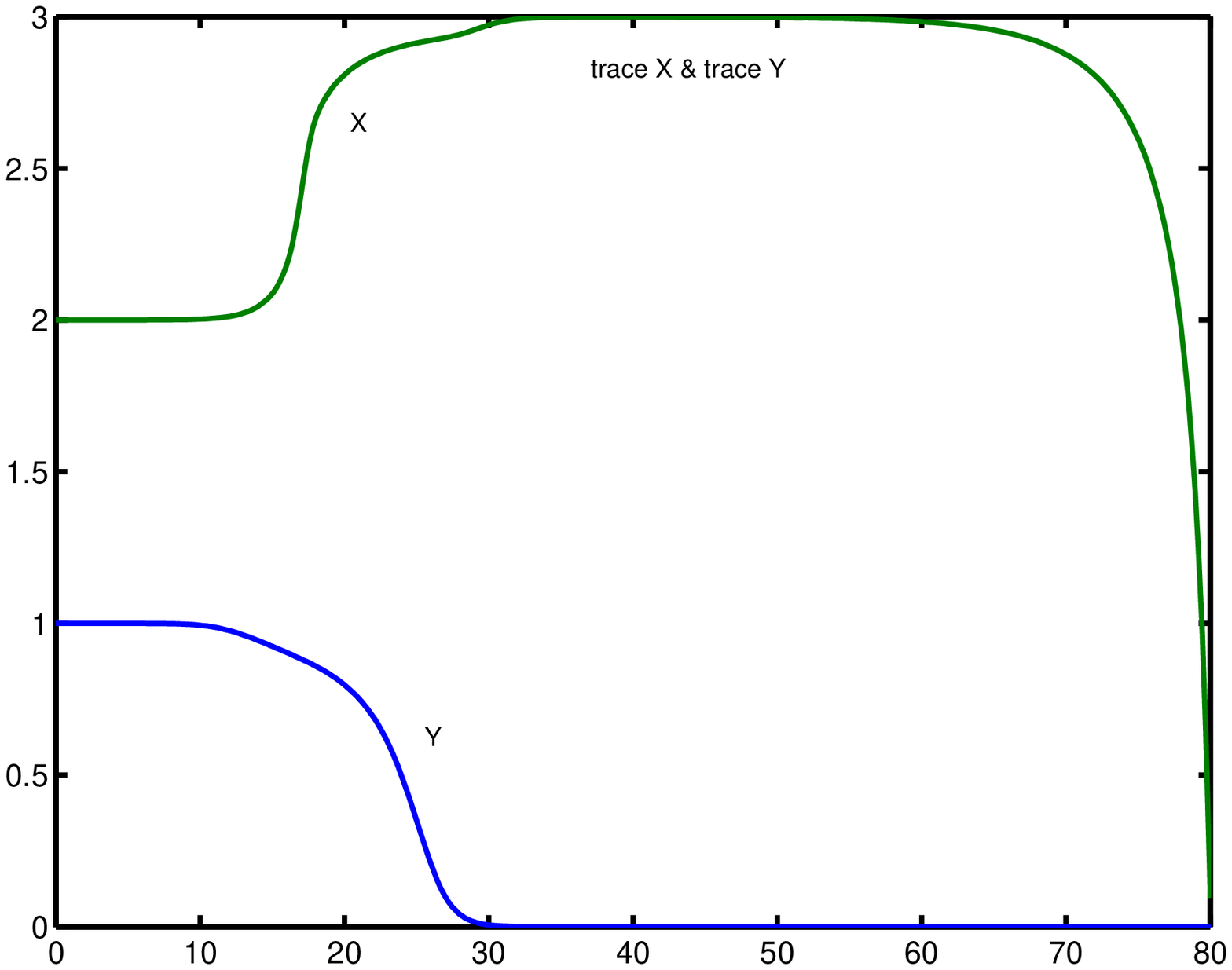,width=6.0cm} \hfill}
\centerline{\hspace*{3.5cm} a \hfill b \hspace*{3.5cm}}
\caption{Behavior of the diagonal elements of matrix $X$ - dashed
line, $Y$- solid line a), traces of the matrices $X$ and $Y$ - b). Here $D_{g}=%
\protect\begin{pmatrix}
-1 & 0.5 \protect \\
0.5 & -1%
\protect\end{pmatrix}%
$, $D_{f}= \protect\begin{pmatrix}
0.5 & -0.15 & 0.5 & 0.4 \protect \\
-0.15 & 1 & 0.3 & 0.1 \protect \\
0.5 & 0.3 & 1 & 0.7 \protect \\
0.4 & 0.1 & 0.7 & 1%
\protect\end{pmatrix}%
$. Initial conditions: $Z(t_0)=%
\protect\begin{pmatrix}
0.05 & -0.02 & 0.0 & 0.0 \protect \\
0 & 0 & 0 & 0%
\protect\end{pmatrix}
$, $x_{1}=(1, 0, 0, 0)^{T}$, $x_{2}=(0,1,0,0)^{T}$, $%
x_{3}=x_{4}=(0,0,0,0)^{T}$ ; $y_{1}=(1,0)^{T}$,
$y_{2}=(0,0)^{T}$.}
\end{figure}

It should be noted that the displayed plot strongly depends on the
parameters of the metrices describing the network evolution. By adjusting
values of the elements of the matrices we can get different network
dynamics. This can correspond to real situation in living organisms when
immune system extinct the antigens or it doesn't and there are many of
natural parameters regulating immune system dynamics.

\begin{figure}[tbph]
\centerline{ \hfill \psfig{figure=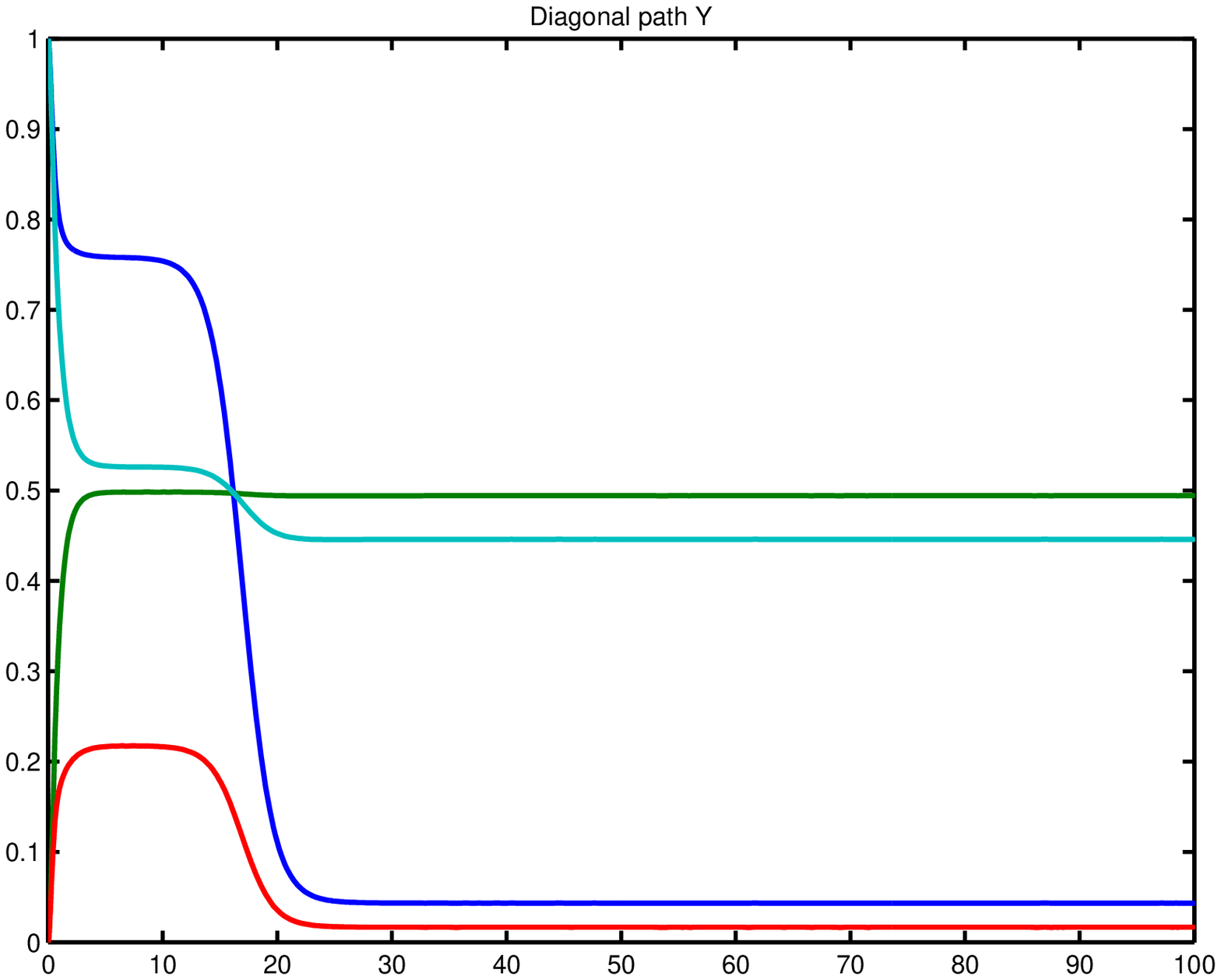,width=6.5cm}
\hfill \psfig{figure=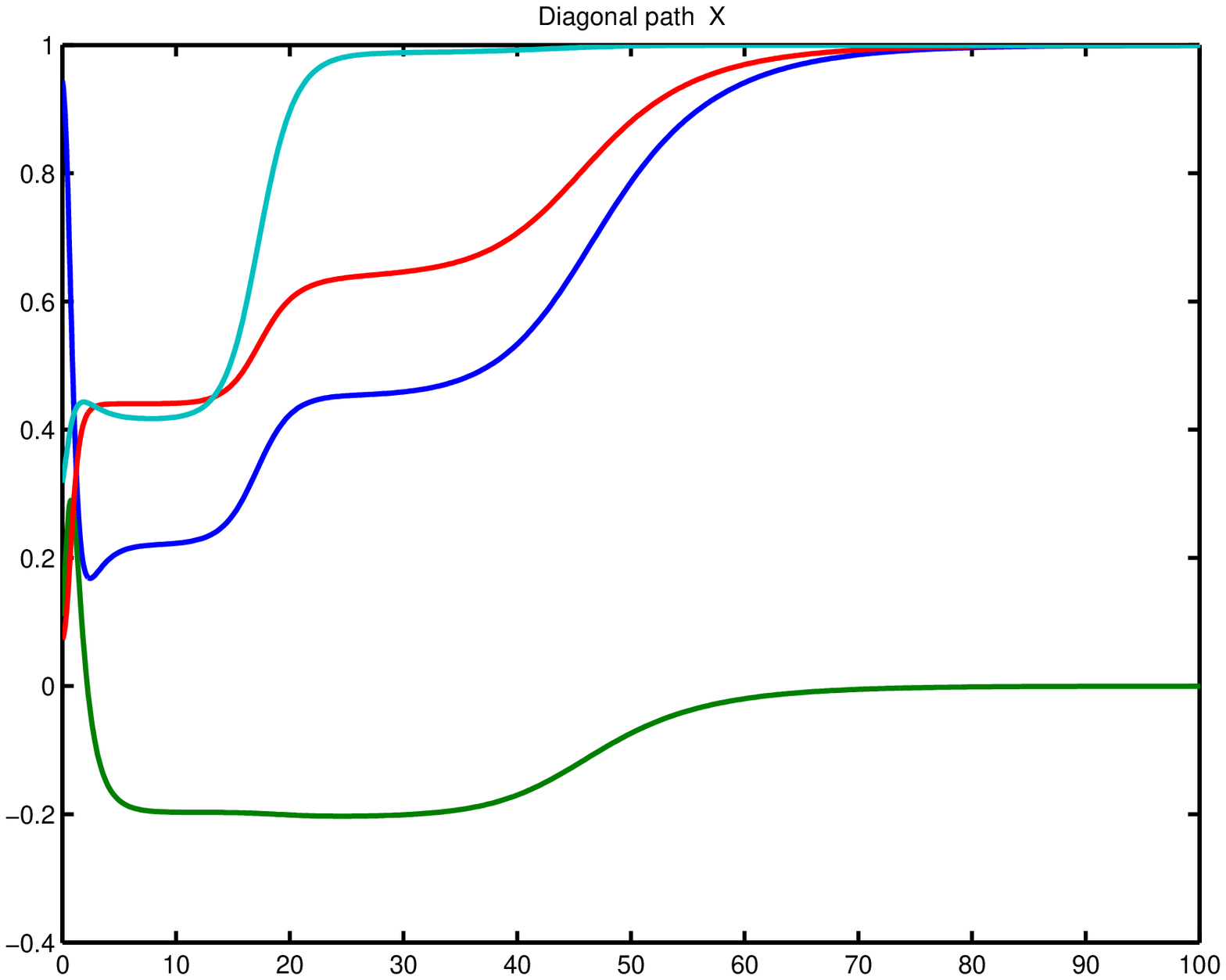,width=6.5cm} \hfill}
\centerline{\hspace*{3.5cm} a \hfill b
\hspace*{3.5cm}}\vspace*{4mm}
\centerline{ \hfill
\psfig{figure=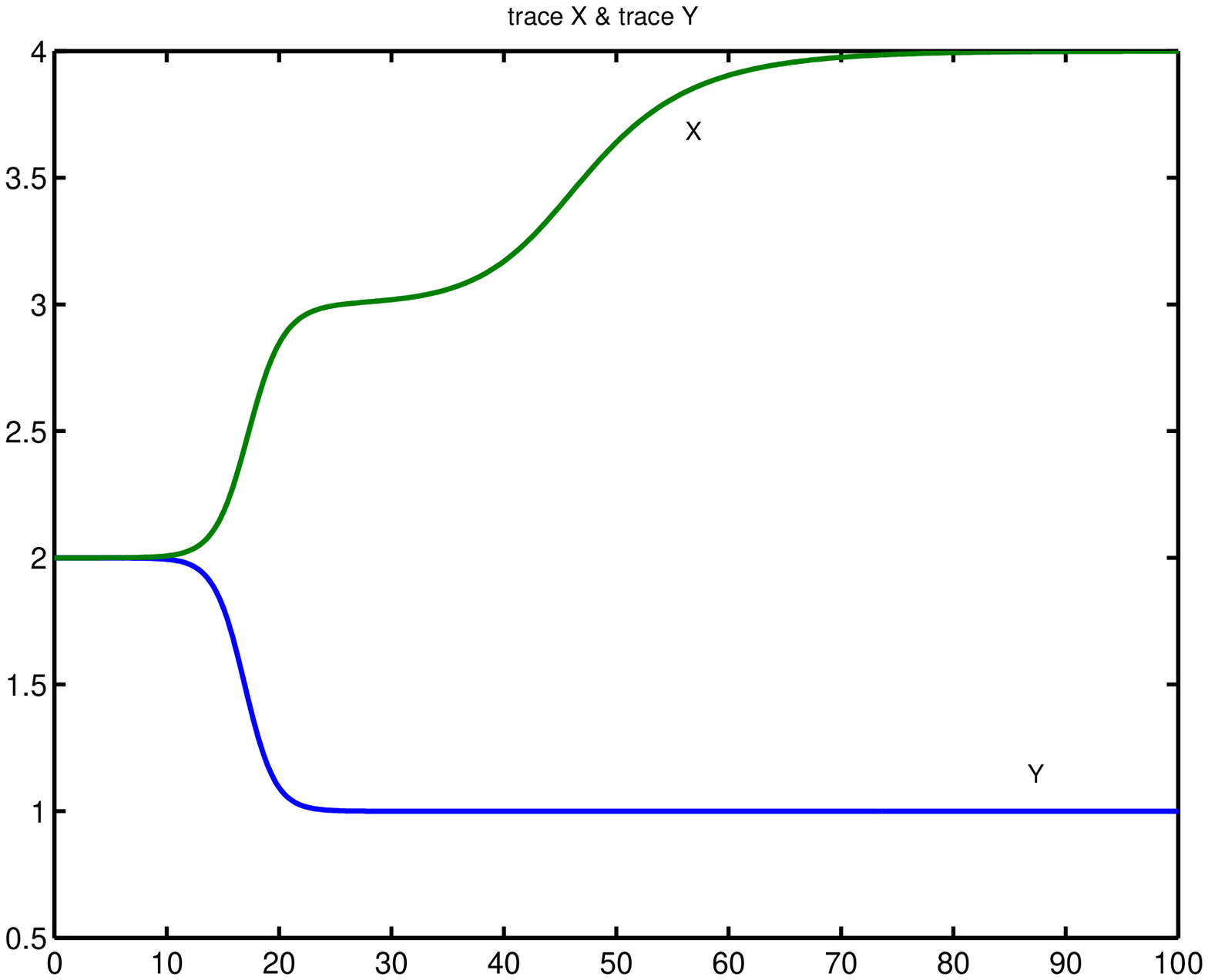,width=6.5cm} \hfill
\psfig{figure=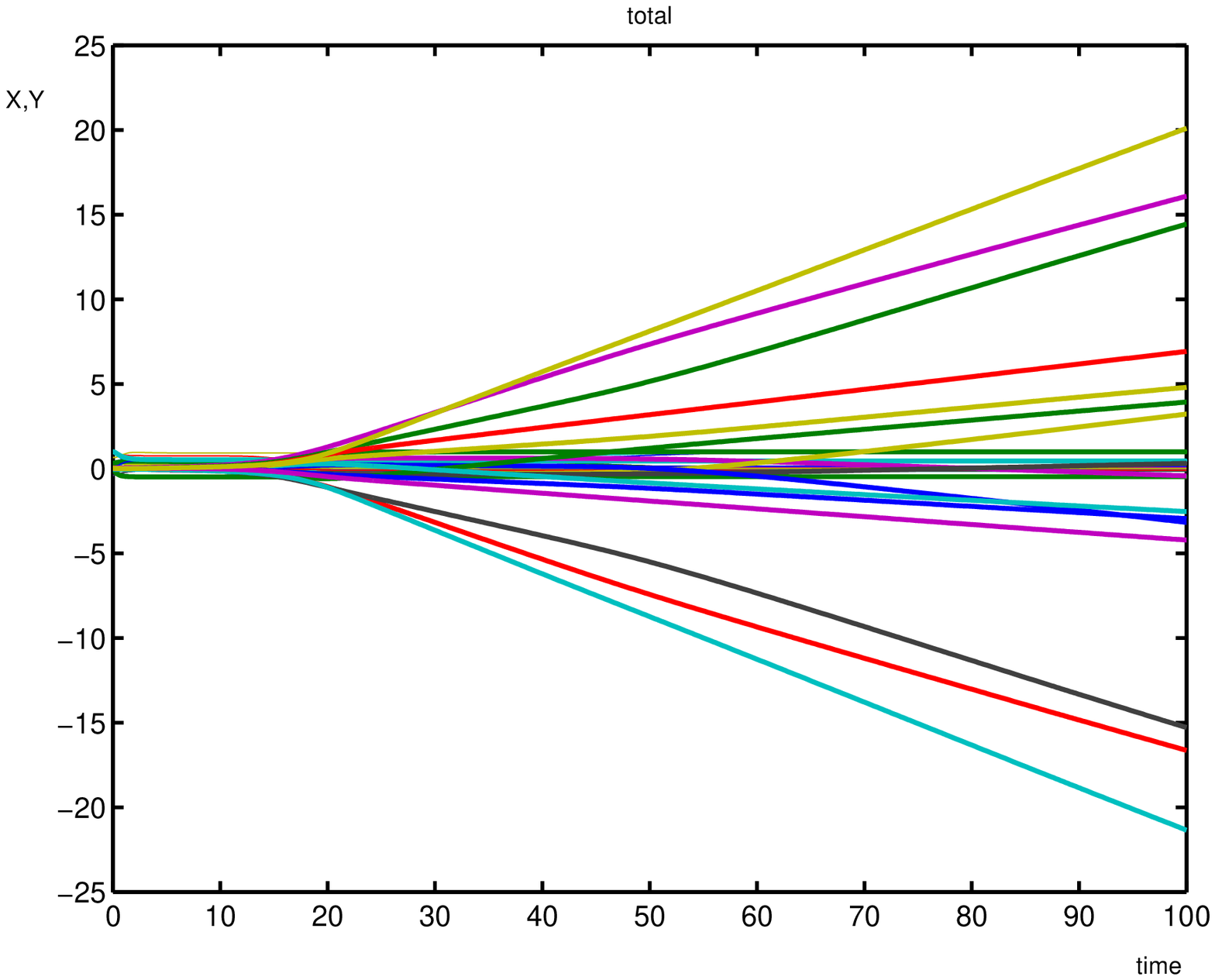,width=6.5cm} \hfill} \centerline{%
\hspace*{3.5cm} c \hfill d \hspace*{3.5cm}}
\caption{Behavior of the diagonal elements of matrix $Y$ - a), $X$ - b),
traces of the matrices $X$ and $Y$ - c), and plot of all solutions of
matrices $X$, $Y$ and $Z$ - d). Here $D_{g}=-%
\protect\begin{pmatrix}
1 & 0.5 & 1 & 0.4 \protect \\
0.5 & 1 & 0.4 & 1 \protect \\
1 & 0.4 & 2 & 0.5 \protect \\
0.4 & 1 & 0.5 & 1%
\protect\end{pmatrix}%
$, $D_{f}=
\protect\begin{pmatrix}
0.5 & -0.1 & 0.5 & 0.4 \protect \\
-0.1 & 1 & 0.3 & 0.1 \protect \\
0.5 & 0.3 & 1 & 0.7 \protect \\
0.4 & 0.1 & 0.7 & a%
\protect\end{pmatrix}%
$, Initial conditions: $Z(t_0)=%
\protect\begin{pmatrix}
0.0522 & 0.0092 & -0.0040 & 0.0087 \protect \\
0 & 0 & 0 & 0 \protect \\
0 & 0 & 0 & 0 \protect \\
0.0097 & -0.0269 & 0.0086 & 0.0202%
\protect\end{pmatrix}
$, $X(t_0)=
\protect\begin{pmatrix}
0.9448 & 0.1090 & -0.0534 & 0.1937 \protect \\
0.1090 & 0.6655 & -0.2193 & -0.4033 \protect \\
-0.0534 & -0.2193 & 0.0726 & 0.1280 \protect \\
0.1937 & -0.4033 & 0.1280 & 0.3172
\protect\end{pmatrix}
$, $x_{1}=(0.9262, -0.1382, 0.0276, 0.3497)^{T}$, $%
x_{2}=(0.2948,0.8040,-0.2680,-0.4415)^{T}$, $x_{3}=(0,0,0,0)^{T}$, $%
x_{4}=(0,0,0,0)^{T}$; $y_{1}=(1,0,0,0)^{T}$, $y_{2}=y_{3}=(0,0,0,0)^{T}$, $%
y_{4}=(0,0,0,1)^{T}$. }
\end{figure}

\begin{figure}[tbph]
\centerline{ \hfill \psfig{figure=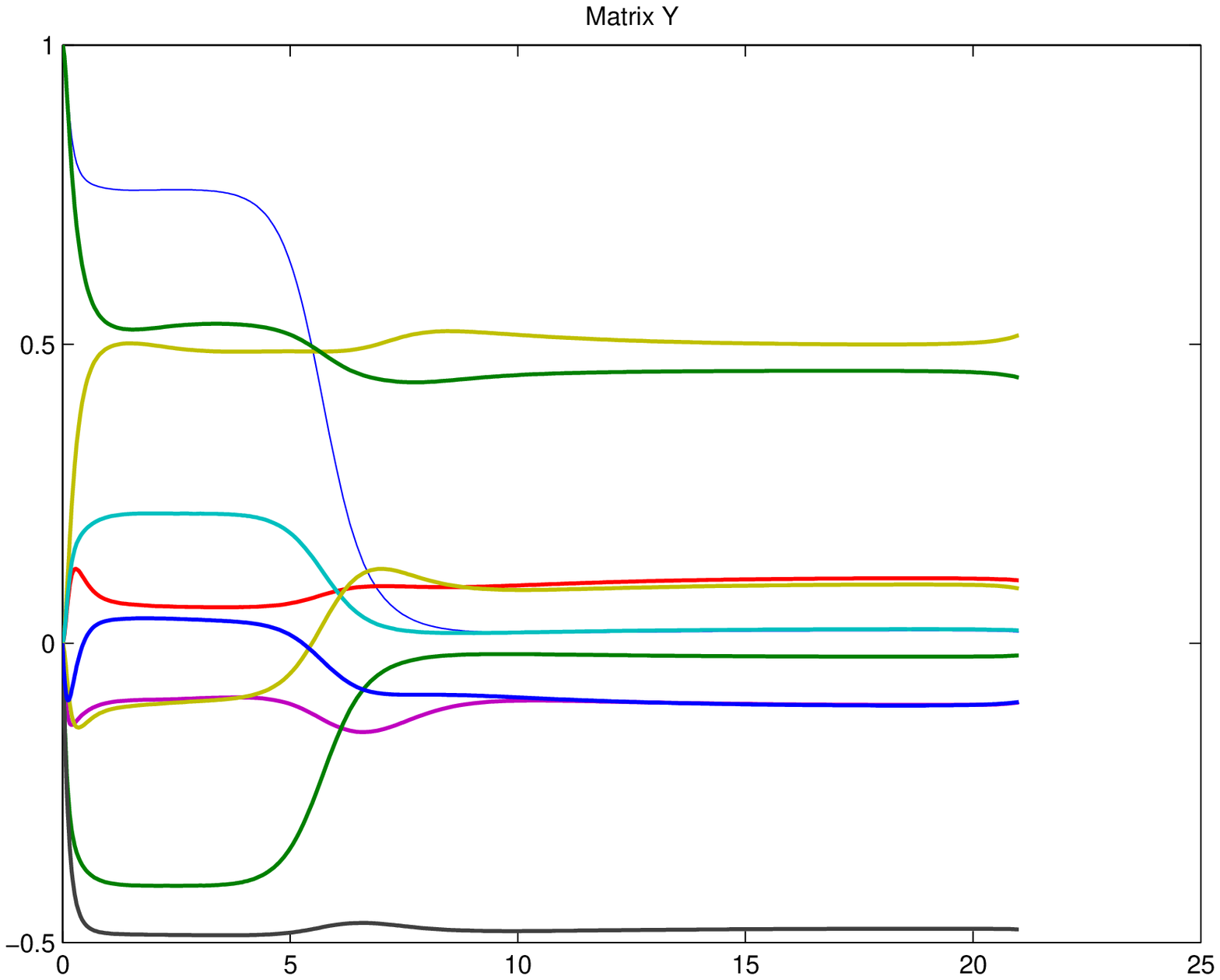,width=6.5cm}
\hfill \psfig{figure=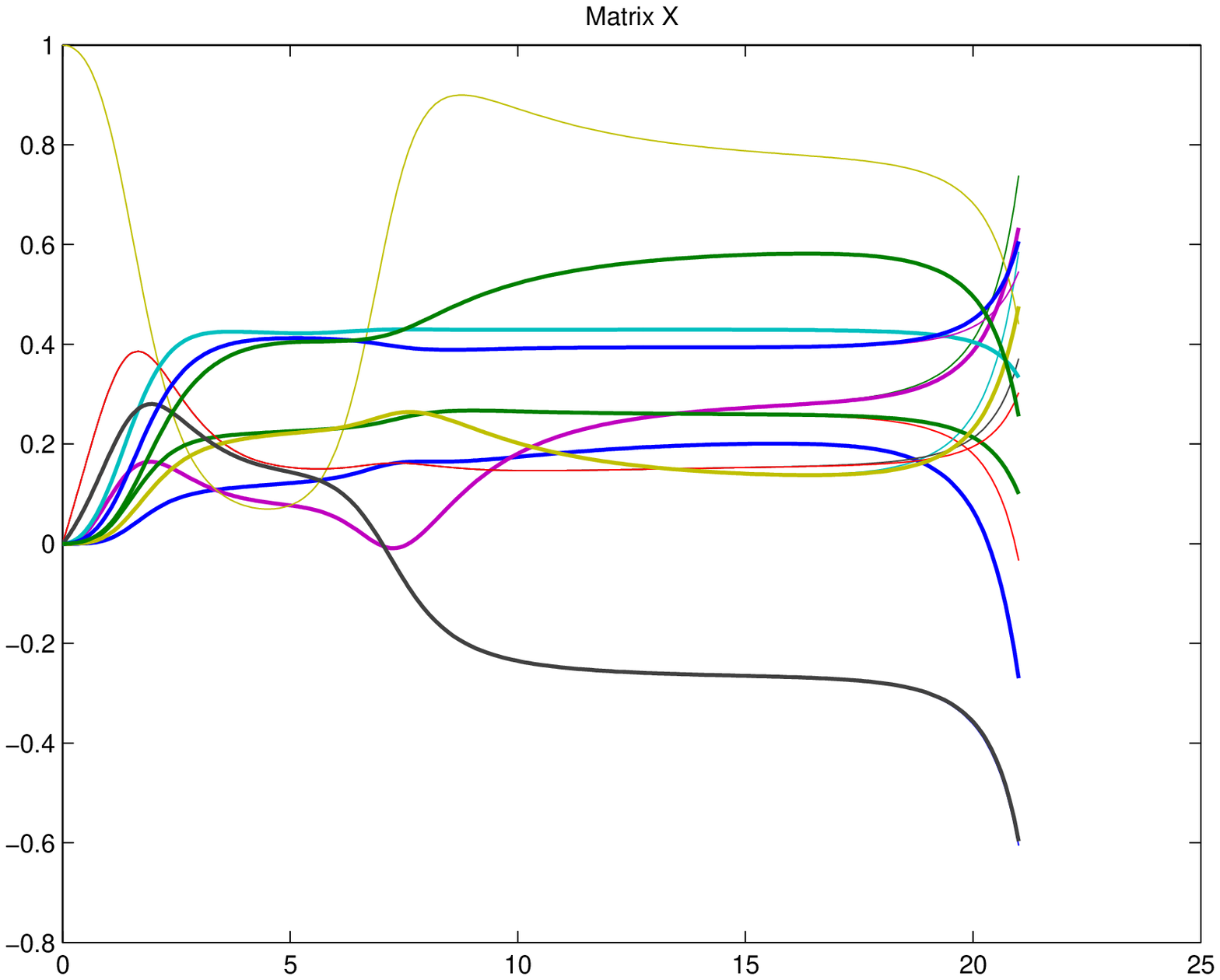,width=6.5cm} \hfill}
\centerline{\hspace*{3.5cm} a \hfill b
\hspace*{3.5cm}}\vspace*{4mm}
\centerline{ \hfill
\psfig{figure=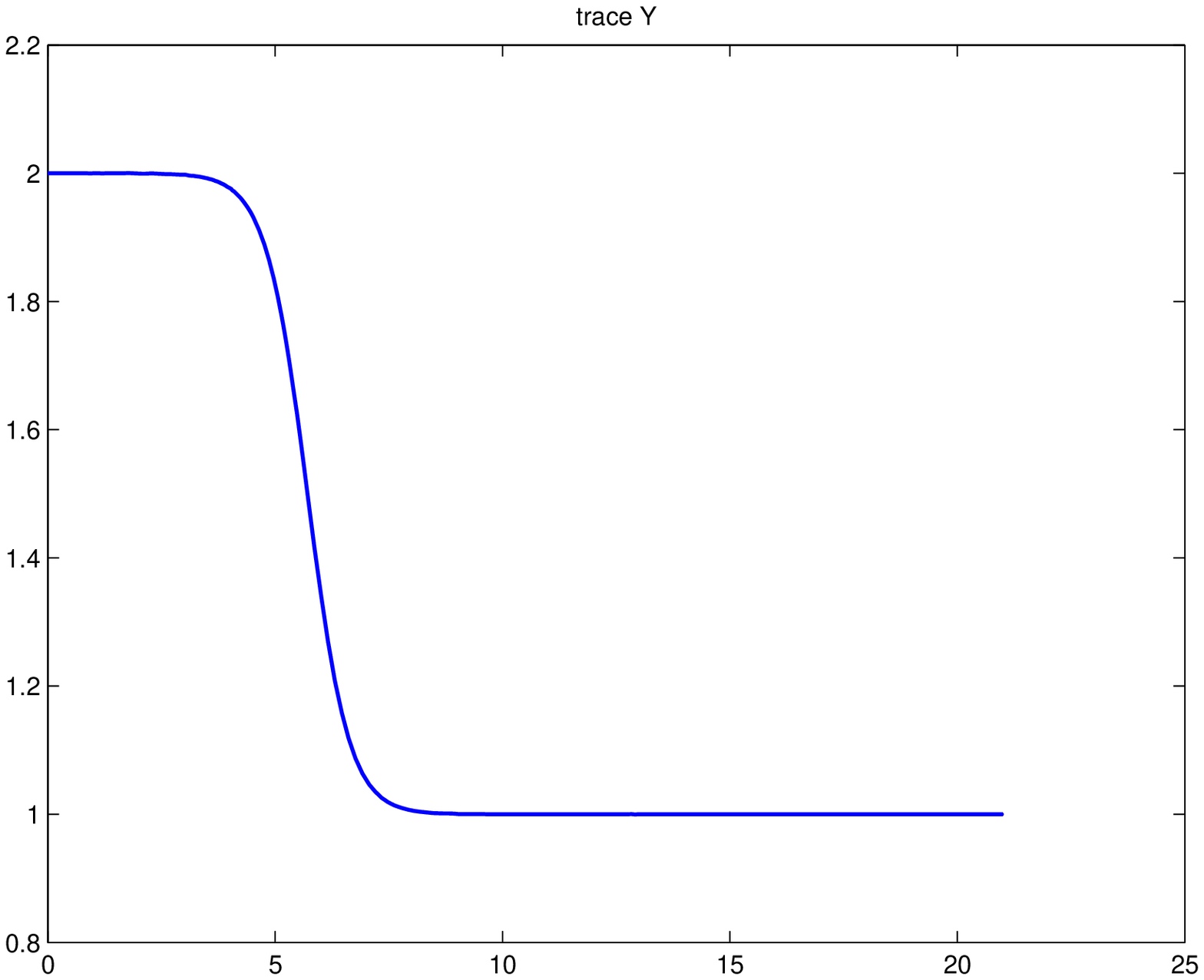,width=6.5cm} \hfill
\psfig{figure=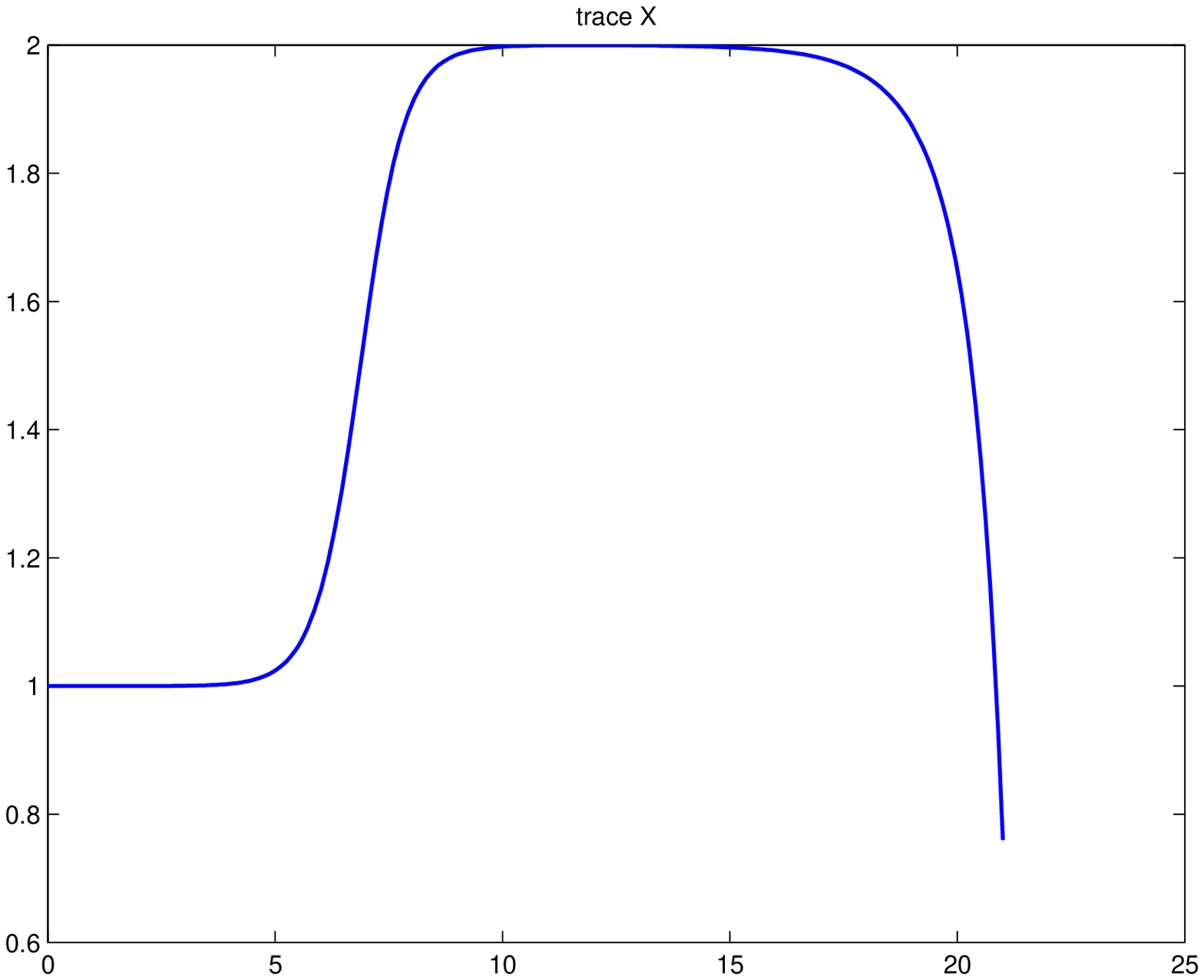,width=6.5cm} \hfill} \centerline{%
\hspace*{3.5cm} c \hfill d \hspace*{3.5cm}}
\caption{ Clonal immune system dynamics for matrix Y - a), for matrix X -
b), Trace of the matrices X and Y - c), d correspondingly. Here $D_{g}=-3%
\protect\begin{pmatrix}
1 & 0.5 & 1 & 0.4 \protect \\
0.5 & 1 & 0.4 & 1 \protect \\
1 & 0.4 & 2 & 0.5 \protect \\
0.4 & 1 & 0.5 & 1%
\protect\end{pmatrix}%
$, $D_{f}=%
\protect\begin{pmatrix}
0.5 & 0 & 0.5 & 0.4 \protect \\
0 & 1 & 0.3 & 0.1 \protect \\
0.5 & 0.3 & 1 & 0.7 \protect \\
0.4 & 0.1 & 0.7 & a%
\protect\end{pmatrix}%
$, $D_{h}=-%
\protect\begin{pmatrix}
0.3 & 0.6 & 0.2 & 0.9 \protect \\
0.5 & 1 & 0.03 & 0.4 \protect \\
0.2 & 0.5 & 0.1 & -0.5 \protect \\
-0.5 & 0.03 & 0.04 & 1%
\protect\end{pmatrix}%
$, Initial conditions: $Z(t_0)=%
\protect\begin{pmatrix}
0.0 & 0.2 & 0.0 & 0.0 \protect \\
0.0 & 0.0 & 0.0 & 0.0 \protect \\
0.0 & 0.0 & 0.0 & 0.0 \protect \\
0.0 & -0.3 & 0.0 & 0.0%
\protect\end{pmatrix}%
$, $x_{1}=x_{3}=x_{4}=(0,0,0,0)^{T }$, $x_{2}=(0,1,0,0)^{T }, $ \ $%
y_{1}=(1,0,0,0)^{T }$, $\ y_{2}=(0,1,0,0)^{T },$ $y_{3}=y_{4}=(0,0,0,0)^{T
}. $}
\end{figure}

\begin{figure}[tbph]
\centerline{ \hfill \psfig{figure=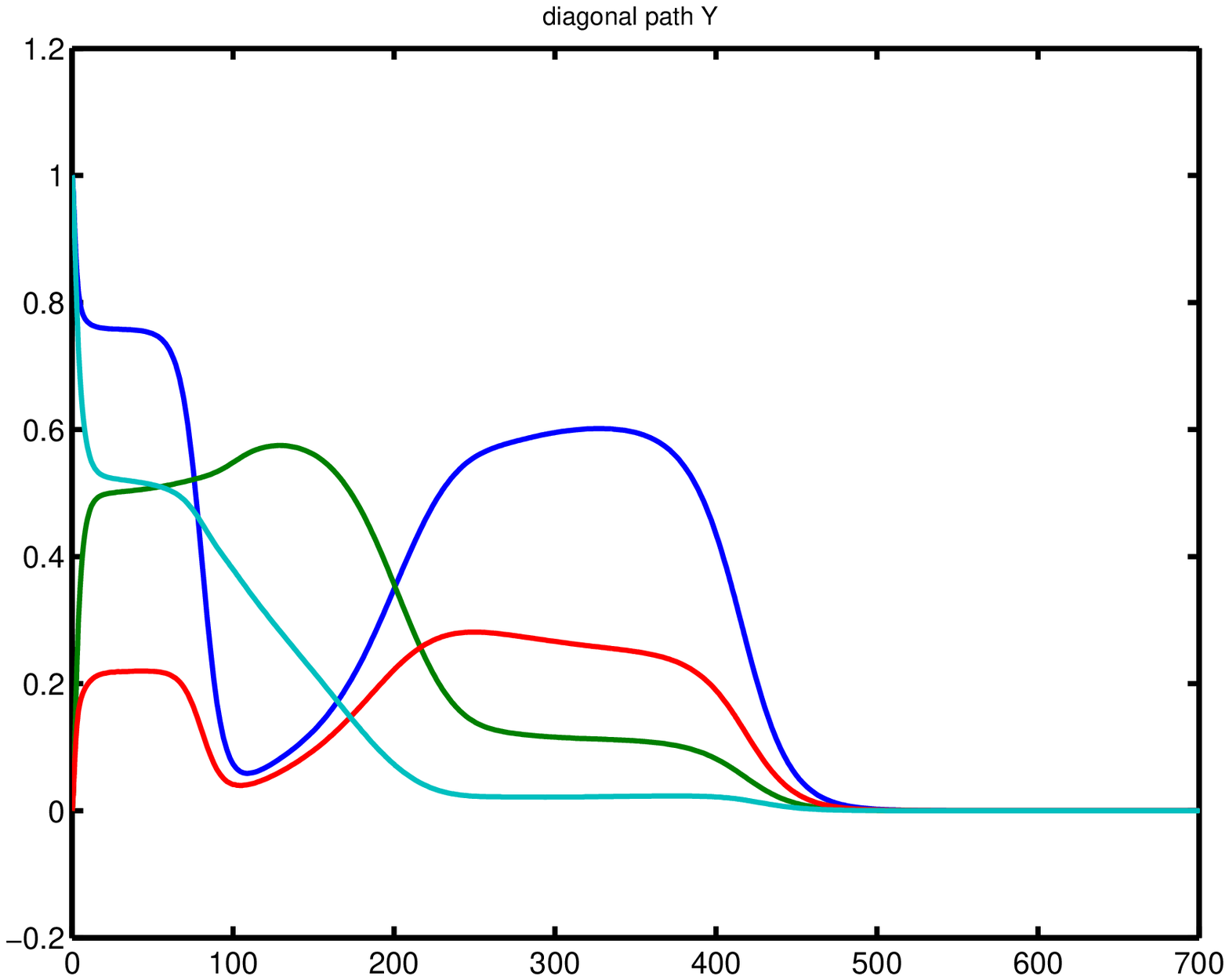,width=6.5cm}
\hfill \psfig{figure=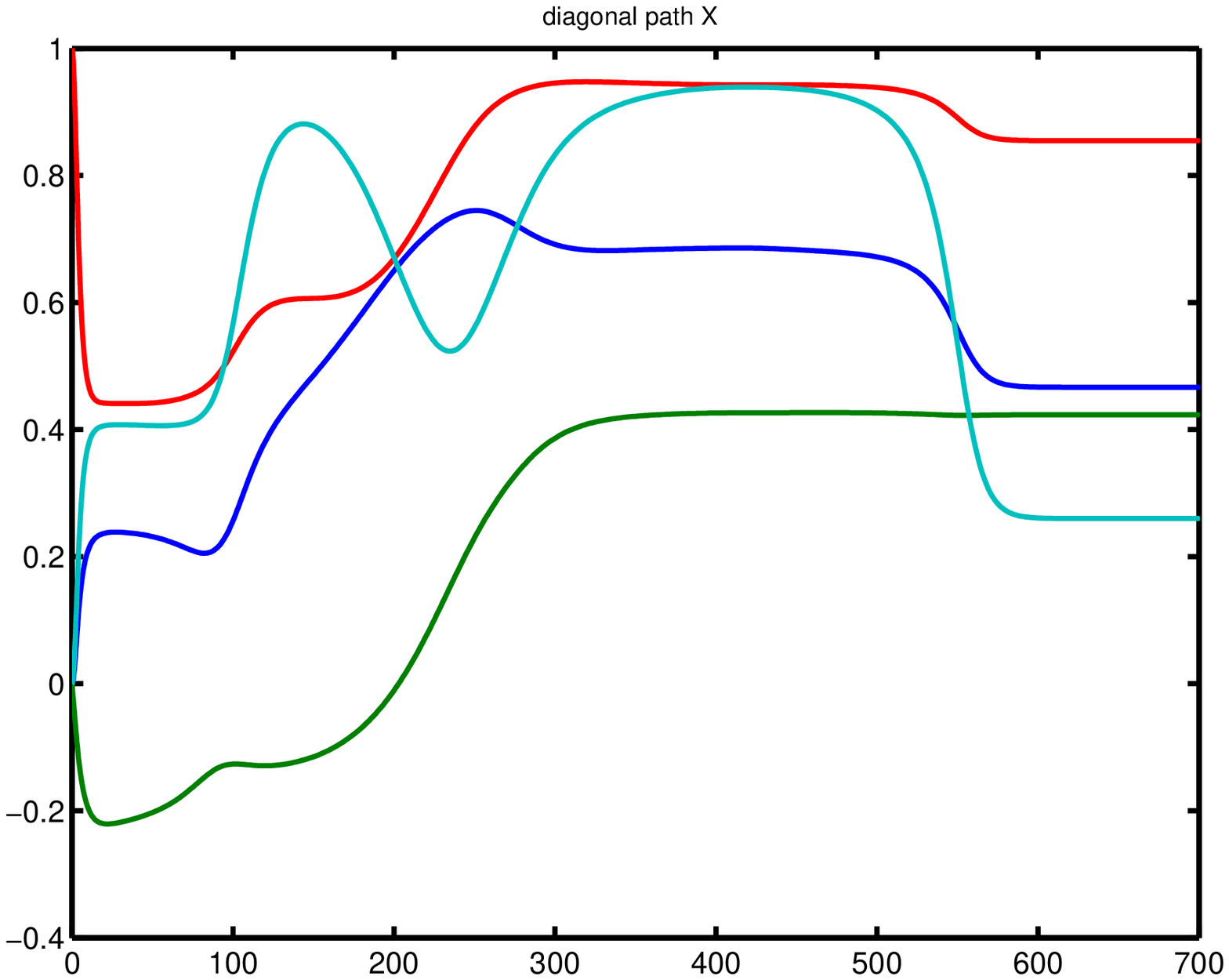,width=6.5cm} \hfill}
\centerline{\hspace*{3.5cm} a \hfill b
\hspace*{3.5cm}}\vspace*{4mm}
\centerline{ \hfill
\psfig{figure=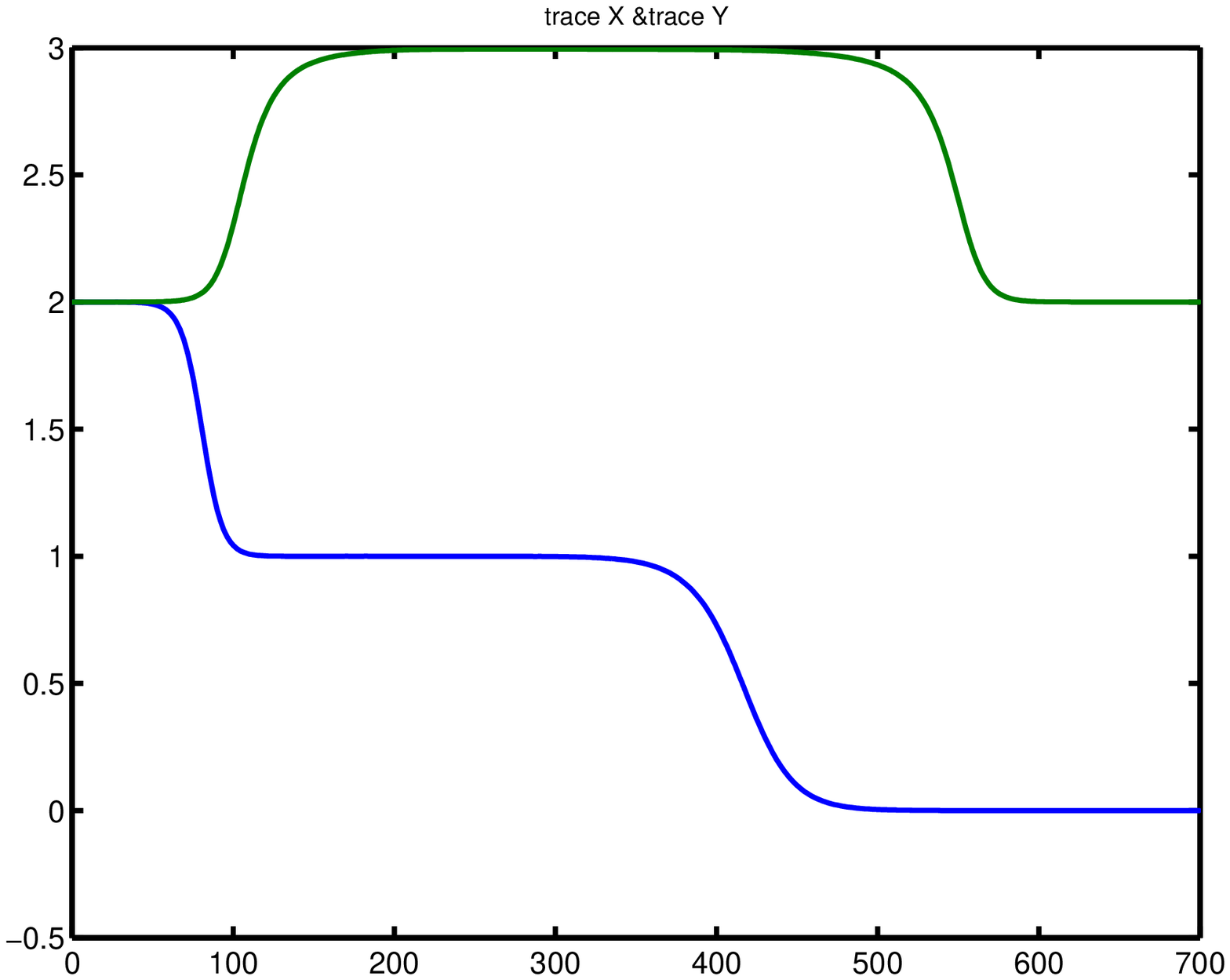,width=6.5cm} \hfill
\psfig{figure=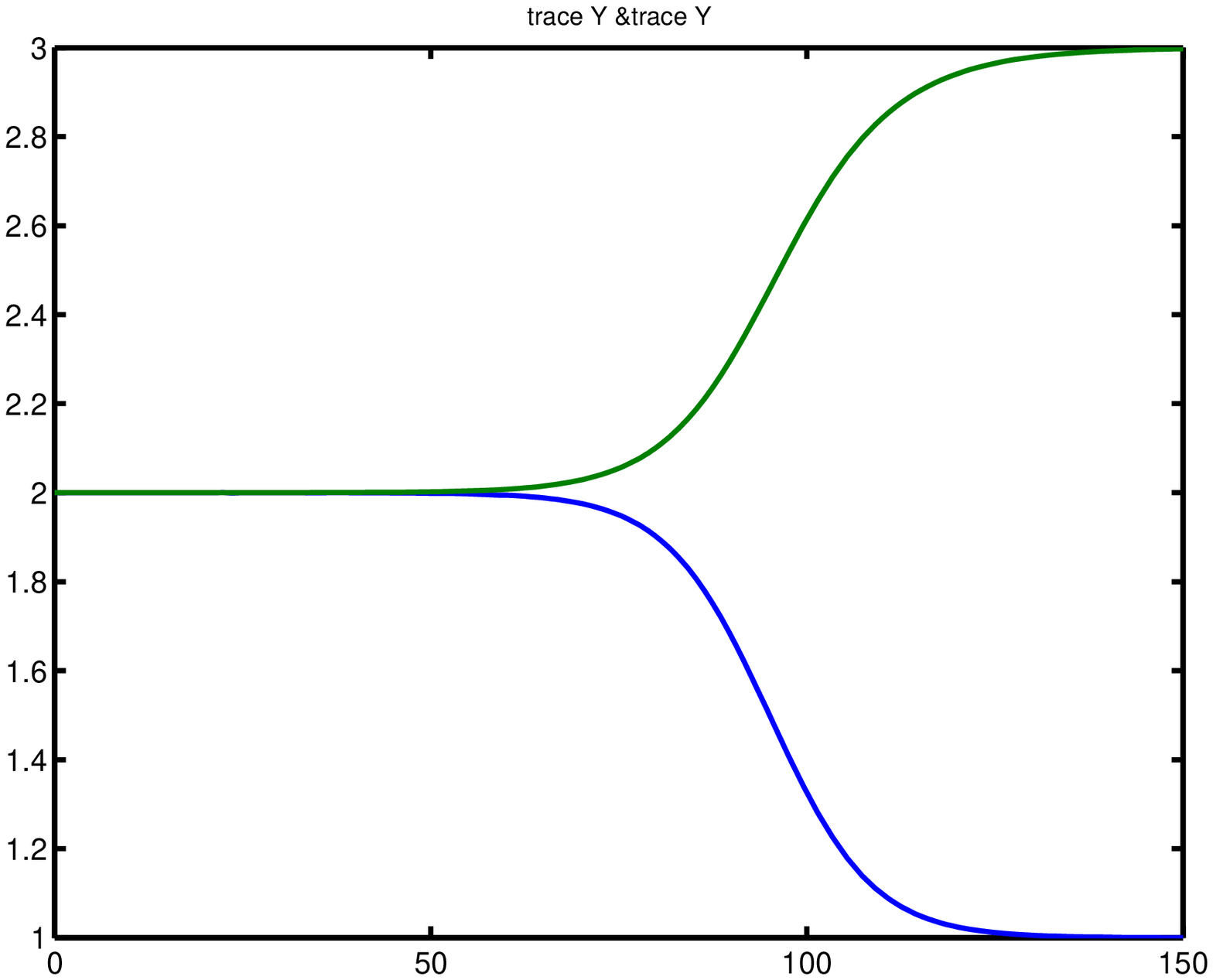,width=6.5cm} \hfill} \centerline{%
\hspace*{3.5cm} c \hfill d \hspace*{3.5cm}}
\caption{ The behavior of the diagonal elements of the matrix $Y$ - a) and $%
X $ - b). Traces of the matrices $X$ and $Y$ for indexes (1) and (2) - c),
d) correspondingly. Here $D_{g}=-0.2%
\protect\begin{pmatrix}
1 & 0.5 & 1 & 0.4 \protect \\
0.5 & 1 & 0.4 & 1 \protect \\
1 & 0.4 & 2 & 0.5 \protect \\
0.4 & 1 & 0.5 & 1%
\protect\end{pmatrix}%
$, $D_{f}=0.2
\protect\begin{pmatrix}
0.5 & 0.15 & 0.5 & 0.4 \protect \\
0.15 & 1 & 0.3 & 0.1 \protect \\
0.5 & 0.3 & 1 & 0.7 \protect \\
0.4 & 0.1 & 0.7 & 1%
\protect\end{pmatrix}%
$, $\protect\beta_{g}=
\protect\begin{pmatrix}
1 & 0.0 & 0.0 & 0.0 \protect \\
0.1 & 1 & 0.1 & 0.0 \protect \\
0.0 & 0.0 & 1 & 0.1 \protect \\
0.0 & 0.0 & 0.0 & 1%
\protect\end{pmatrix}%
$, $\protect\beta_{f}=
\protect\begin{pmatrix}
1 & 0.0 & 0.0 & 0.0 \protect \\
0.1 & 1 & 0.2 & 0.0 \protect \\
0.0 & 0.0 & 1 & 0.1 \protect \\
0.0 & 0.2 & 0.0 & 1%
\protect\end{pmatrix}%
$. Initial conditions: $Z(t_0)=%
\protect\begin{pmatrix}
0.0 & 0.2 & 0.03 & 0.0 \protect \\
0.0 & 0.0 & 0.0 & 0.0 \protect \\
0.0 & 0.0 & 0.0 & 0.0 \protect \\
0.0 & -0.3 & 0.02 & 0.0%
\protect\end{pmatrix}%
$, $x_{1}=x_{4}=(0,0,0,0)^{T }$, $x_{2}=(0,1,0,0)^{T }$, $x_{3}=(0,0,1,0)^{T
}$, $y_{1}=(1,0,0,0)^{T }$, $\ y_{4}=(0,0,0,1)^{T }$, $%
y_{2}=y_{3}=(0,0,0,0)^{T }.$}
\end{figure}

In Fig. 2 we plotted the outcome of numerical simulations for the case $\dim
X={4=\dim }Y.$ As in the previous case, for simplicity, we consider the
matrix $D_{h}$ being zero. The numerical solution of the dynamical system
obtained for certain $D_{g}$ and $D_{f}$ \ also corresponds to the case when
interaction between clonal variables $X$ and $Y$ is completely determined by
matrices $D_{g}$ and $D_{f}$ and, namely, they change the adaptive
interaction of the main matrix variables through the matrix variable $Z$.
Examination of the solutions obtained reveals that dynamics of the diagonal
elements of the matrices $X$ and $Y\in End\mathbb{R}^{4}$ - a) and b) and
traces of these matrices c) is essentially nonlinear. After some
quasi-stationary dynamics of the traces we have their sharp flips to new
integer values. During the system evolution a new increase or decrease of $X$
-components can happen which may change essentially its trace dynamics. For
initial conditions we used two dimensional subspace of the matrices $X$ and $%
Y$. Concerning inhibitor $X$ we chose arbitrary orthogonal vectors $%
x_{1,}x_{2}$ and concerning activator vectors we chose two orthonormal
vectors. These vectors form matrices $X,Y$ as well at some given values of
coefficients $z_{\alpha \beta }$ do the matrix $Z$. The specific data of
these procedure are presented in caption for Figure 2. From the reported
plot we can see that nondiagonal phase trajectories (Fig. 2, d) of the
matrices $X$, $Y$ and $Z$ have practically linear behavior and are
relatively stable for large values of time $(t\sim 1000)$.

In Figure 3 a) we report the numerical simulation of all components of
matrices $X$- a) and $Y$- b) and corresponding trace dynamics of these
matrices $X$- c) and $Y$- d) for the case with new initial data and all
matrices being nonequal to zero. In particular, we took the rank of the
matrix $X$ \ in initial conditions equal to be constant $1$ and that of the
matrix $Y$ equal to $2$ and initial vectors are standard orthogonal vectors
(See Fig. 3).

The form of the matrix $D_{h}$ plays essential role in interaction of the
clonal network. From the reported plot we can see that phase trajectories of
activator $Y$ are stabilized at certain value of $t\in \mathbb{R}_{+}$ and
the values of $Y$ \ practically doesn't change in time. The simulations run
until the time when the system either destroys owing computational errors or
stabilizes.

In Figure 4 we present the numerical simulation of the diagonal components
of matrices $X$- a) and $Y$- b) and the corresponding trace dynamics of
these matrices -c) for the network dynamics stimulated by combination of two
Lyapunov functions (\ref{2}), (\ref{6}) and (\ref{2}), (\ref{8}). In these
cases new matrices in equations are determined as
\begin{equation*}
\widetilde{D}_{g}=k_{g}D_{g}+D_{g}^{(1,2)},~\widetilde{D}%
_{f}=k_{f}D_{f}+D_{f}^{(1,2)},\widetilde{D}_{h}=k_{h}D_{h}+D_{h}^{(1,2)}
\end{equation*}%
but equations of the clonal dynamics are persist evidently to be the same as
(\ref{ad1}). The first upper index (1) corresponds to Lyapunov function (\ref%
{6}) and the second upper index (2) corresponds to Lyapunov function (\ref{8}%
). The results of computer simulations for first index are presented on
Figure 4 a-c) and the second one on the Figure 4 - d). In this case the
diagonal elements $X$ - a) diagonal elements of matrix Y - b) and traces of
these two matrices c) demonstrate a very interesting nonlinear dynamics. For
simplicity of representation and in order to grasp some characteristic
features of the clonal network dynamics we used practically the same
matrices as in the previous simulations. All parameters of matrices $D_{f},$
$D_{g},$ $D_{h},$ $D_{f}^{(1)}=D_{f}^{(2)}$ and $D_{g}^{(1)}=D_{g}^{(2)}$
are presented on captions to this figure. With our simulations we studied
the effects of the combination influence of these two Lyapunov potentials.

It was established within these simulations that coefficients before the
medication matrices influence sufficiently the network dynamics. It should
be noted that the behavior of the system with Lyapunov potential
combinations of these two potentials (\ref{2}) and (\ref{6}), (\ref{8}) are
much more stabilized and we have many opportunities by taking coefficients $%
k_{g},k_{f},k_{h}$ small enough to obtain different trace dynamics of the
system. Analyzing behavior of the traces describing our immune clonal
network we can conclude that small variations of the parameters do not
change seriously the clonal dynamics. In this case there exists some region
of attraction when the solutions persist the same form. But at some values
of elements of matrices these solutions change drastically. This is the case
happening in the real immune system dynamics when inhibitory receptors
possess many of parameters which behave precisely in order to depress
antigens from the clonal network.

Within our model of an immune clonal system the matrix trace $trY\in \mathbb{%
Z}_{+}$ of the matrix $Y,$ being an integer, counts the number of activated
excitatory receptors strings at the moment of time $t\in \mathbb{R}_{+}$
during the interaction of excitatory clonal sub-network with inhibitory
clonal sub-network, described at the same moment by the matrix trace $trX\in
\mathbb{Z}_{+}$ of the matrix $X,$ being also an integer and counting,
respectively, the number of activated inhibitory receptors strings during
the interaction. As one can see, during the interaction between inhibitory
and excitatory clones at some moments of time there are switched some new
excitatory and inhibitory receptors strings, and further at next whiles of
time during their interactions, some of excitatory receptors strings become
depressed and some inhibitory receptors either appear or become
dis-activated too. This event can be interpreted, for instance, as follows:
our inhibitory clonal network solved its immune task to dis-activate the
excitatory clonal sub-network, and next at some later while of time becomes
idle too, leaving itself in the initial awaiting state.

Thereby, one can state that the model studied in this paper possesses many
of properties suitable for possible responses of a real clonal network. The
stimulation of the interaction between excitatory and inhibitory sub-systems
demonstrates their expectable direct self-similar and complementary
behavior. This model is obviously not still completely satisfactory because
it needs many of external parameters accounting for the main important
features of real clonal network dynamics. But we believe that the work
presented in this paper is an alternative step towards better understanding
the essence and nature of immune network dynamics. Since the realistic
immune networks do involve much more than several receptor strings elements,
such an analysis is a new step to a completely novel approach to
understanding the functioning of real clonal immune networks.

\end{document}